\documentclass[aps,prab,superscriptaddress,preprint]{revtex4-2}
\usepackage{natbib}
\usepackage[utf8]{inputenc}
\usepackage{graphicx}
\usepackage{siunitx}
\usepackage{subfigure}
\usepackage{booktabs}
\usepackage{url}
\usepackage{libertine,libertinust1math}
\usepackage[T1]{fontenc}

\usepackage{lineno}
\let\oldalign\align
\let\oldendalign\endalign
\renewcommand{\align}{\linenomath\oldalign}
\renewcommand{\endalign}{\oldendalign\endlinenomath}

\newif\ifdraft
\draftfalse 
\ifdraft
\linenumbers
\fi

\begin{document}
\title{Measurement of the emittance of accelerated electron bunches at the AWAKE experiment}

\author{D.~A.~Cooke}
\email{Contact author: david.cooke@ucl.ac.uk}
\affiliation{University College London, Gower St., London, WC1E 6BT, United Kingdom}
\author{F.~Pannell}
\affiliation{University College London, Gower St., London, WC1E 6BT, United Kingdom}
\author{G.~Zevi Della Porta}
\affiliation{CERN, 1211 Geneva 23, Switzerland}
\affiliation{Max Planck Institute for Physics, 80805 Munich, Germany}
\author{J.~Farmer}
\affiliation{Max Planck Institute for Physics, 80805 Munich, Germany}
\author{V.~Bencini}
\affiliation{CERN, 1211 Geneva 23, Switzerland}
\affiliation{John Adams Institute, Oxford University, Oxford OX1 3RH, United Kingdom}
\author{M.~Bergamaschi}
\affiliation{Max Planck Institute for Physics, 80805 Munich, Germany}
\author{S.~Mazzoni}
\affiliation{CERN, 1211 Geneva 23, Switzerland}
\author{L.~Ranc}
\affiliation{Max Planck Institute for Physics, 80805 Munich, Germany}
\author{E.~Senes}
\affiliation{CERN, 1211 Geneva 23, Switzerland}
\author{P.~Sherwood}
\affiliation{University College London, Gower St., London, WC1E 6BT, United Kingdom}
\author{M.~Wing}
\affiliation{University College London, Gower St., London, WC1E 6BT, United Kingdom}
\collaboration{AWAKE Collaboration}
\noaffiliation
\author{R.~Agnello}
\affiliation{Ecole Polytechnique Federale de Lausanne (EPFL), Swiss Plasma Center (SPC), 1015 Lausanne, Switzerland}
\author{C.C.~Ahdida}
\affiliation{CERN, 1211 Geneva 23, Switzerland}
\author{C.~Amoedo}
\affiliation{CERN, 1211 Geneva 23, Switzerland}
\author{Y.~Andrebe}
\affiliation{Ecole Polytechnique Federale de Lausanne (EPFL), Swiss Plasma Center (SPC), 1015 Lausanne, Switzerland}
\author{O.~Apsimon}
\affiliation{University of Manchester M13 9PL, Manchester M13 9PL, United Kingdom}
\affiliation{Cockcroft Institute, Warrington WA4 4AD, United Kingdom}
\author{R.~Apsimon}
\affiliation{Cockcroft Institute, Warrington WA4 4AD, United Kingdom} 
\affiliation{Lancaster University, Lancaster LA1 4YB, United Kingdom}
\author{J.M.~Arnesano}
\affiliation{CERN, 1211 Geneva 23, Switzerland}
\author{P.~Blanchard}
\affiliation{Ecole Polytechnique Federale de Lausanne (EPFL), Swiss Plasma Center (SPC), 1015 Lausanne, Switzerland}
\author{P.N.~Burrows}
\affiliation{John Adams Institute, Oxford University, Oxford OX1 3RH, United Kingdom}
\author{B.~Buttensch{\"o}n}
\affiliation{Max Planck Institute for Plasma Physics, 17491 Greifswald, Germany}
\author{A.~Caldwell}
\affiliation{Max Planck Institute for Physics, 80805 Munich, Germany}
\author{M.~Chung}
\affiliation{POSTECH, Pohang 37673, Republic of Korea}
\author{A.~Clairembaud}
\affiliation{CERN, 1211 Geneva 23, Switzerland}
\author{C.~Davut}
\affiliation{University of Manchester M13 9PL, Manchester M13 9PL, United Kingdom}
\affiliation{Cockcroft Institute, Warrington WA4 4AD, United Kingdom} 
\author{G.~Demeter}
\affiliation{HUN-REN Wigner Research Centre for Physics, Budapest, Hungary}
\author{A.C.~Dexter}
\affiliation{Cockcroft Institute, Warrington WA4 4AD, United Kingdom} 
\affiliation{Lancaster University, Lancaster LA1 4YB, United Kingdom}
\author{S.~Doebert}
\affiliation{CERN, 1211 Geneva 23, Switzerland}
\author{A.~Fasoli}
\affiliation{Ecole Polytechnique Federale de Lausanne (EPFL), Swiss Plasma Center (SPC), 1015 Lausanne, Switzerland}
\author{R.~Fonseca}
\affiliation{ISCTE - Instituto Universit\'{e}ario de Lisboa, 1049-001 Lisbon, Portugal}  
\affiliation{GoLP/Instituto de Plasmas e Fus\~{a}o Nuclear, Instituto Superior T\'{e}cnico, Universidade de Lisboa, 1049-001 Lisbon, Portugal}
\author{I.~Furno}
\affiliation{Ecole Polytechnique Federale de Lausanne (EPFL), Swiss Plasma Center (SPC), 1015 Lausanne, Switzerland}
\author{N.Z.~van~Gils}
\affiliation{CERN, 1211 Geneva 23, Switzerland}
\affiliation{PARTREC, UMCG, University of Groningen, Groningen, The Netherlands}
\author{E.~Granados}
\affiliation{CERN, 1211 Geneva 23, Switzerland}
\author{M.~Granetzny}
\affiliation{University of Wisconsin, Madison, WI 53706, USA}
\author{T.~Graubner}
\affiliation{Philipps-Universit{\"a}t Marburg, 35032 Marburg, Germany}
\author{O.~Grulke}
\affiliation{Max Planck Institute for Plasma Physics, 17491 Greifswald, Germany}
\affiliation{Technical University of Denmark, 2800 Kgs. Lyngby, Denmark}
\author{E.~Gschwendtner}
\affiliation{CERN, 1211 Geneva 23, Switzerland}
\author{E.~Guran}
\affiliation{CERN, 1211 Geneva 23, Switzerland}
\author{J.~Henderson}
\affiliation{Cockcroft Institute, Warrington WA4 4AD, United Kingdom}
\affiliation{STFC/ASTeC, Daresbury Laboratory, Warrington WA4 4AD, United Kingdom}
\author{M.Á.~Kedves}
\affiliation{HUN-REN Wigner Research Centre for Physics, Budapest, Hungary}
\author{F.~Kraus}
\affiliation{Philipps-Universit{\"a}t Marburg, 35032 Marburg, Germany}
\author{M.~Krupa}
\affiliation{CERN, 1211 Geneva 23, Switzerland}
\author{T.~Lefevre}
\affiliation{CERN, 1211 Geneva 23, Switzerland}
\author{L.~Liang}
\affiliation{University of Manchester M13 9PL, Manchester M13 9PL, United Kingdom}
\affiliation{Cockcroft Institute, Warrington WA4 4AD, United Kingdom}
\author{S.~Liu}
\affiliation{TRIUMF, Vancouver, BC V6T 2A3, Canada}
\author{N.~Lopes}
\affiliation{GoLP/Instituto de Plasmas e Fus\~{a}o Nuclear, Instituto Superior T\'{e}cnico, Universidade de Lisboa, 1049-001 Lisbon, Portugal}
\author{K.~Lotov}
\affiliation{Budker Institute of Nuclear Physics SB RAS, 630090 Novosibirsk, Russia}
\affiliation{Novosibirsk State University, 630090 Novosibirsk , Russia}
\author{M.~Martinez~Calderon}
\affiliation{CERN, 1211 Geneva 23, Switzerland}
\author{J.~Mezger}
\affiliation{Max Planck Institute for Physics, 80805 Munich, Germany}
\author{P.I.~Morales~Guzm\'{a}n}
\affiliation{Max Planck Institute for Physics, 80805 Munich, Germany}
\author{M.~Moreira}
\affiliation{GoLP/Instituto de Plasmas e Fus\~{a}o Nuclear, Instituto Superior T\'{e}cnico, Universidade de Lisboa, 1049-001 Lisbon, Portugal}
\author{T.~Nechaeva}
\affiliation{Max Planck Institute for Physics, 80805 Munich, Germany}
\author{N.~Okhotnikov}
\affiliation{Budker Institute of Nuclear Physics SB RAS, 630090 Novosibirsk, Russia}
\affiliation{Novosibirsk State University, 630090 Novosibirsk , Russia}
\author{C.~Pakuza}
\affiliation{John Adams Institute, Oxford University, Oxford OX1 3RH, United Kingdom}
\author{A.~Pardons}
\affiliation{CERN, 1211 Geneva 23, Switzerland}
\author{K.~Pepitone}
\affiliation{Angstrom Laboratory, Department of Physics and Astronomy, 752 37 Uppsala, Sweden}
\author{E.~Poimendidou}
\affiliation{CERN, 1211 Geneva 23, Switzerland}
\author{J.~Pucek}
\affiliation{Max Planck Institute for Physics, 80805 Munich, Germany}
\author{A.~Pukhov}
\affiliation{Heinrich-Heine-Universit{\"a}t D{\"u}sseldorf, 40225 D{\"u}sseldorf, Germany}
\author{R.L.~Ramjiawan}
\affiliation{CERN, 1211 Geneva 23, Switzerland}
\affiliation{John Adams Institute, Oxford University, Oxford OX1 3RH, United Kingdom}
\author{S.~Rey}
\affiliation{CERN, 1211 Geneva 23, Switzerland}
\author{R.~Rossel}
\affiliation{CERN, 1211 Geneva 23, Switzerland}
\author{H.~Saberi}
\affiliation{University of Manchester M13 9PL, Manchester M13 9PL, United Kingdom}
\affiliation{Cockcroft Institute, Warrington WA4 4AD, United Kingdom}
\author{O.~Schmitz}
\affiliation{University of Wisconsin, Madison, WI 53706, USA}
\author{F.~Silva}
\affiliation{INESC-ID, Instituto Superior Técnico, Universidade de Lisboa, 1049-001 Lisbon, Portugal}
\author{L.~Silva}
\affiliation{GoLP/Instituto de Plasmas e Fus\~{a}o Nuclear, Instituto Superior T\'{e}cnico, Universidade de Lisboa, 1049-001 Lisbon, Portugal}
\author{B.~Spear}
\affiliation{John Adams Institute, Oxford University, Oxford OX1 3RH, United Kingdom}
\author{C.~Stollberg}
\affiliation{Ecole Polytechnique Federale de Lausanne (EPFL), Swiss Plasma Center (SPC), 1015 Lausanne, Switzerland}
\author{A.~Sublet}
\affiliation{CERN, 1211 Geneva 23, Switzerland}
\author{C.~Swain}
\affiliation{Cockcroft Institute, Warrington WA4 4AD, United Kingdom}
\affiliation{University of Liverpool, Liverpool L69 7ZE, United Kingdom}
\author{A.~Topaloudis}
\affiliation{CERN, 1211 Geneva 23, Switzerland}
\author{N.~Torrado}
\affiliation{CERN, 1211 Geneva 23, Switzerland}
\affiliation{GoLP/Instituto de Plasmas e Fus\~{a}o Nuclear, Instituto Superior T\'{e}cnico, Universidade de Lisboa, 1049-001 Lisbon, Portugal}
\author{P.~Tuev}
\affiliation{Budker Institute of Nuclear Physics SB RAS, 630090 Novosibirsk, Russia}
\affiliation{Novosibirsk State University, 630090 Novosibirsk , Russia}
\author{F.~Velotti}
\affiliation{CERN, 1211 Geneva 23, Switzerland}
\author{V.~Verzilov}
\affiliation{TRIUMF, Vancouver, BC V6T 2A3, Canada}
\author{J.~Vieira}
\affiliation{GoLP/Instituto de Plasmas e Fus\~{a}o Nuclear, Instituto Superior T\'{e}cnico, Universidade de Lisboa, 1049-001 Lisbon, Portugal}
\author{E.~Walter}
\affiliation{Max Planck Institute for Plasma Physics, Munich, Germany}
\author{C.~Welsch}
\affiliation{Cockcroft Institute, Warrington WA4 4AD, United Kingdom}
\affiliation{University of Liverpool, Liverpool L69 7ZE, United Kingdom}
\author{M.~Wendt}
\affiliation{CERN, 1211 Geneva 23, Switzerland}
\author{J.~Wolfenden}
\affiliation{Cockcroft Institute, Warrington WA4 4AD, United Kingdom}
\affiliation{University of Liverpool, Liverpool L69 7ZE, United Kingdom}
\author{B.~Woolley}
\affiliation{CERN, 1211 Geneva 23, Switzerland}
\author{G.~Xia}
\affiliation{Cockcroft Institute, Warrington WA4 4AD, United Kingdom}
\affiliation{University of Manchester M13 9PL, Manchester M13 9PL, United Kingdom}
\author{L.~Verra}
\altaffiliation{Present Address: INFN Laboratori Nazionali di Frascati, 00044 Frascati, Italy}
\affiliation{CERN, 1211 Geneva 23, Switzerland}
\author{V.~Yarygova}
\affiliation{Budker Institute of Nuclear Physics SB RAS, 630090 Novosibirsk, Russia}
\affiliation{Novosibirsk State University, 630090 Novosibirsk , Russia}
\author{M.~Zepp}
\affiliation{University of Wisconsin, Madison, WI 53706, USA}

\begin{abstract}
	The vertical plane transverse emittance of accelerated electron bunches at the AWAKE experiment at CERN has been determined, using three different methods of data analysis. This is a proof-of-principle measurement using the existing AWAKE electron spectrometer to validate the measurement technique. Large values of the geometric emittance, compared to that of the injection beam, are observed ($\sim \SI{0.5}{\milli\metre\milli\radian}$ compared with $\sim \SI{0.08}{\milli\metre\milli\radian}$), which is in line with expectations of emittance growth arising from plasma density ramps and large injection beam bunch size. Future iterations of AWAKE are anticipated to operate in conditions where emittance growth is better controlled, and the effects of the imaging systems of the existing and future spectrometer designs on the ability to measure the emittance are discussed. Good performance of the instrument down to geometric emittances of approximately $\SI{1e-4}{\milli\metre\milli\radian}$ is required, which may be possible with improved electron optics and imaging.
\end{abstract}

\maketitle

\section{Introduction}

The Advanced WAKEfield (AWAKE) experiment at CERN is a proof-of-concept, particle-driven plasma wakefield accelerator \cite{Gschwendtner2016}. It uses long (\SI{\sim 6}{\centi\metre}) proton bunches delivered by the Super Proton Synchrotron at CERN at \SI{400}{\giga\electronvolt} to drive wakefields in a \SI{10}{\metre} Rb plasma column. The plasma is created using a \SI{120}{\femto\second} laser pulse passing through Rb vapour in a temperature-controlled chamber with Rb vapour sources at either end \cite{Plyushchev2018}. The laser and proton bunches propagate coaxially and cotemporally; the use of a long proton bunch relies on the self-modulation instability \cite{Adli2019} to effectively drive large wakefields, and the plasma density step created by the laser ionization front provides the seed for this instability, ensuring phase repeatability \cite{Batsch2021}. This instability leads to periodic focussing and defocussing of the proton bunch, eventually resulting in a train of microbunches separated by one plasma wavelength (that is, $\frac{2\pi c}{\omega_p}$, where $\frac{c}{\omega_p}$ is the plasma skin depth, $\omega_p$ being the electron plasma frequency). Electrons are injected into the wakefield from a \SI{18}{\mega\electronvolt} combined photoinjector and S-band booster \cite{Pepitone2016,Kim2020}, the laser incident on the photocathode being derived from the main ionizing laser, which ensures phase stability with the seeded self-modulation instability. This in turn ensures that the electrons are, in principle, injected into the same phase of the wakefield each time. In practice, the electron bunch length is larger than the microbunch separation, has a transverse size on the scale of the plasma skin depth, and is subject to jitters in e.g. pointing, so different parts of the bunch (and successive shots) experience different parts of the wakefield. Using this injection scheme, acceleration up to \SI{2}{\giga\electronvolt} of part of the injected bunch has been observed \cite{Adli2018} for a plasma density of \SI{7e14}{\per\centi\metre\cubed}.\par

Normalized emittance is an invariant quantity under acceleration of a particle bunch, given in a single plane by:
\begin{align}
	\epsilon_n &= \gamma\beta\epsilon\\
	\epsilon_n &= \gamma\beta\sqrt[]{\det\Sigma}
\end{align}
where $\epsilon$ is the geometric emittance (phase space projection area), $\gamma$ and $\beta$ the usual Lorentz factors, and $\Sigma$, the beam matrix, is
\begin{align}
	\Sigma &= \left[
	\begin{matrix}
		\sigma_x^2 & \sigma_{xx'} \\ \sigma_{xx'} & \sigma_{y'}^2
	\end{matrix} \right].
\end{align}
Single-shot measurement of the emittance at AWAKE is desirable, as the variation in the injection process could lead to variation in the shot-to-shot accelerated beam parameters. Measurement techniques which rely on aggregated data from multiple shots will therefore not necessarily produce the correct result. As the proton drive beam and accelerated witness electrons propagate coaxially on exit from the plasma cell, measurements such as pepperpot emittance meters or analysis of optical transition radiation are impossible because the signal from the proton beam will vastly dominate. This leaves beam image analysis in a spectrometer as the only currently viable technique for determining the beam parameters of a single accelerated bunch. Although emittance is not expected to be preserved or well controlled with the present experimental geometry, measuring it is both interesting to test whether changes in the acceleration conditions can be detected through changes in the beam parameters, and a way to develop the instrumentation required for later AWAKE runs (e.g. run 2c).

\section{Experimental setup}

\begin{figure}
	\includegraphics[width=\textwidth]{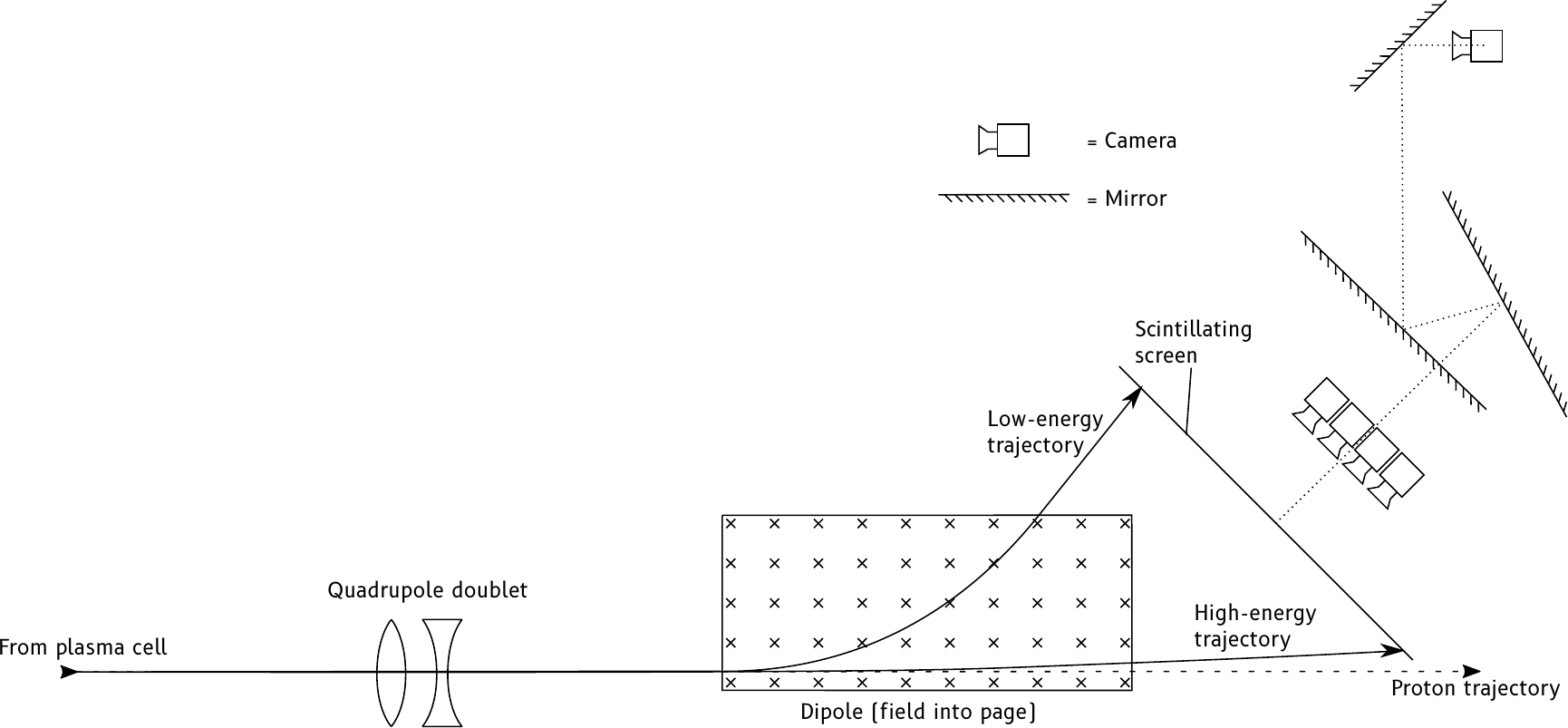}
	\caption{Diagram of the AWAKE electron spectrometer (not to scale).}
	\label{fig:exp}
\end{figure}

Accelerated electron bunches are detected and characterized using a wide-energy-acceptance spectrometer, comprising a quadrupole doublet followed by a single dipole bend. On exit from the dipole, the beam passes through a thin aluminium vacuum window, and a \SI{1}{\metre} wide Gadox scintillating screen, which is viewed by an array of cameras positioned at a distance of \SI{1.25}{\metre} and out of the dipole bending plane at an angle of 30$\degree$. It is also viewed by a single, intensified camera, at a distance of \SI{17}{\metre} which, in this instance, is used to provide calibrated energy and charge measurements. Further details of this spectrometer can be found in \cite{Bauche2019,Keeble2019}, a diagram showing the layout is shown in Figure \ref{fig:exp}.\par

A doublet configuration with identical strength quadrupoles cannot focus or image in both planes at the same distance. The spectrometer is intended to operate as an imaging device, reproducing the object plane of the exit iris of the plasma cell at the image plane of the scintillating screen. The doublet therefore has a resistive shunt across one magnet, which enables vertical and horizontal imaging at a fixed distance. The distance between the final quadrupole and the detection plane varies by approximately \SI{30}{\centi\metre} (the path length difference between the high-energy and low-energy trajectories indicated in Figure \ref{fig:exp}) however, which means this imaging condition is only met at a single point on the scintillating screen. This means that vertical beam size measurement methods for determining emittance in the vertical plane should account for defocussing in the horizontal plane. In practice, this effect can be ignored at the measurement precision acheived in this work.\par

The magnification of the system is approximately 1, so features in the vertical direction at the imaging plane are the same size as at the object plane. The smallest resolvable feature size is determined by a number of factors: electron optics, electron scattering in the aluminium vacuum window, optical scattering of scintillation light in the screen, and the point spread function (PSF) of the camera and lens combination used to image the screen. These factors have been studied, providing an estimate of the overall PSF of the spectrometer.\par

\section{Extraction of Emittance}

\begin{table}
	\centering
	\caption{Parameter values describing the working point of the experiment.}
	\label{tab:wp}
	\begin{tabular}{r|l}
		\hline\hline
		Parameter & Value\\\hline
		Proton bunch population	&	$1\times10^{11}$\\
		Plasma density			&	$1\times10^{14}$ cm$^{-3}$\\
		Laser--proton delay		&	$-30$ ps (on-axis)\\
								&	$+20$ ps (off-axis)\\
		Injection--laser delay	&	$-360$ ps\\
		Injection angle			&	0 mrad (on-axis)\\
								&	1 mrad (off-axis)\\
		Injection focus			&	0 m (on-axis)\\
								&	$+1$ m (off-axis)\\
		Injection bunch charge	&	600 pC\\
		Injection emittance		&	0.08 mm mrad\\\hline\hline
	\end{tabular}
\end{table}

The data for this emittance study were acquired during a single day of operation of the AWAKE experiment. The working point is summarized in Table \ref{tab:wp}; two slightly different trajectory sets are used, differing by injection angle only, with one set being injected approximately coaxially with the laser and proton beams, and the other injected at approximately \SI{1}{\milli\radian} (and so correspondingly offset from the other beams), aimed to cross the axis at \SI{1}{\metre} into the plasma. These sets are hereafter referred to as \textit{on-axis} and \textit{off-axis} respectively. These datasets were acquired as two multi-shot emittance measurements using the standard quadrupole scan technique \cite[e.g.][]{Wiedemann2015}. This technique, where an upstream quadrupole magnet is varied in strength, and the downstream size is measured, assumes there is no shot-to-shot variation of the beam parameters---not necessarily a safe assumption given possible variations in the injection position on the scale of the plasma skin depth. Under such circumstances, a single-shot method for determining the emittance is preferable, and fortunately the spectrometer allows such a measurement to be made, in the non-dispersing plane. This method \cite{Deacon2016,Barber2018} exploits the dispersion by energy in the horizontal plane resulting in each vertical slice having experienced a different quadrupole strength, meaning a single spectrometer image is, in effect, an image of a quadrupole scan. This is slightly complicated by the emittance in the dispersive plane, and the fact that the imaging condition is not necessarily met in this plane, even when it is in the vertical plane. These two effects lead to blurring of the vertical profile. The effects of emittance in the horizontal plane have been studied using simulation, however, and is largely insignificant at the beam parameter scale observed here---assuming that the beam parameters are similar in both horizontal and vertical planes ($x$ and $y$, respectively). This allows the beam size to be written in terms of the vertical plane transport and beam matrices, and the beam matrix elements to be extracted by least-squares fitting this curve to multiple observations of the beam size. Note the single-shot method has different caveats to the multishot quadrupole scan---each energy slice is taken as representative of the whole bunch, that is, the beam parameters do not vary with energy. \par

\begin{figure}
	\centering
	\includegraphics[width=0.49\textwidth]{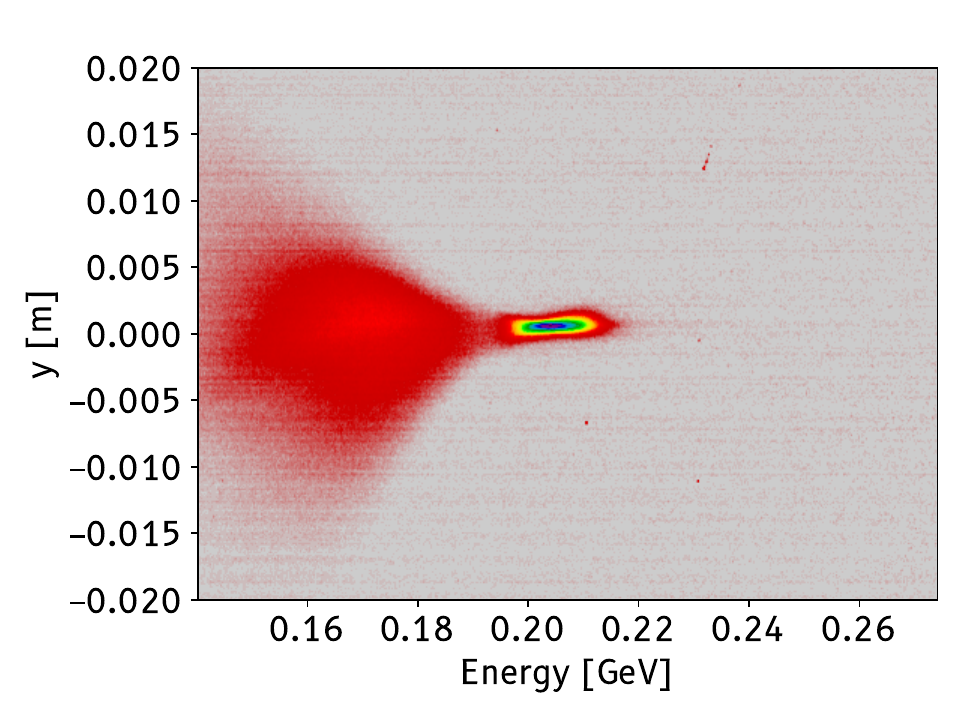}
	\includegraphics[width=0.49\textwidth]{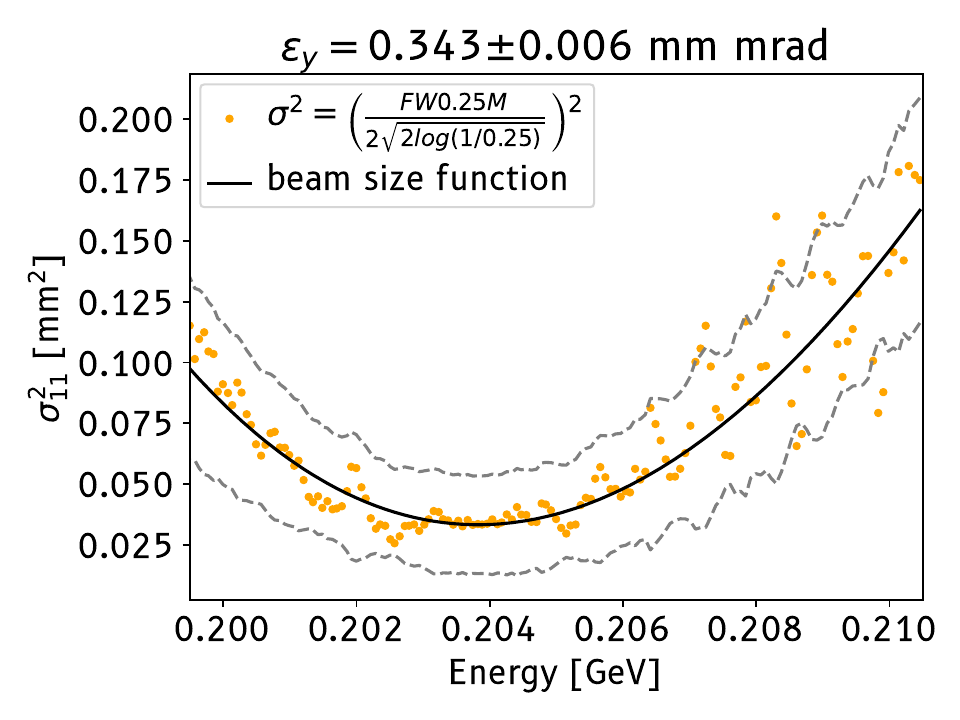}
	\includegraphics[width=0.49\textwidth]{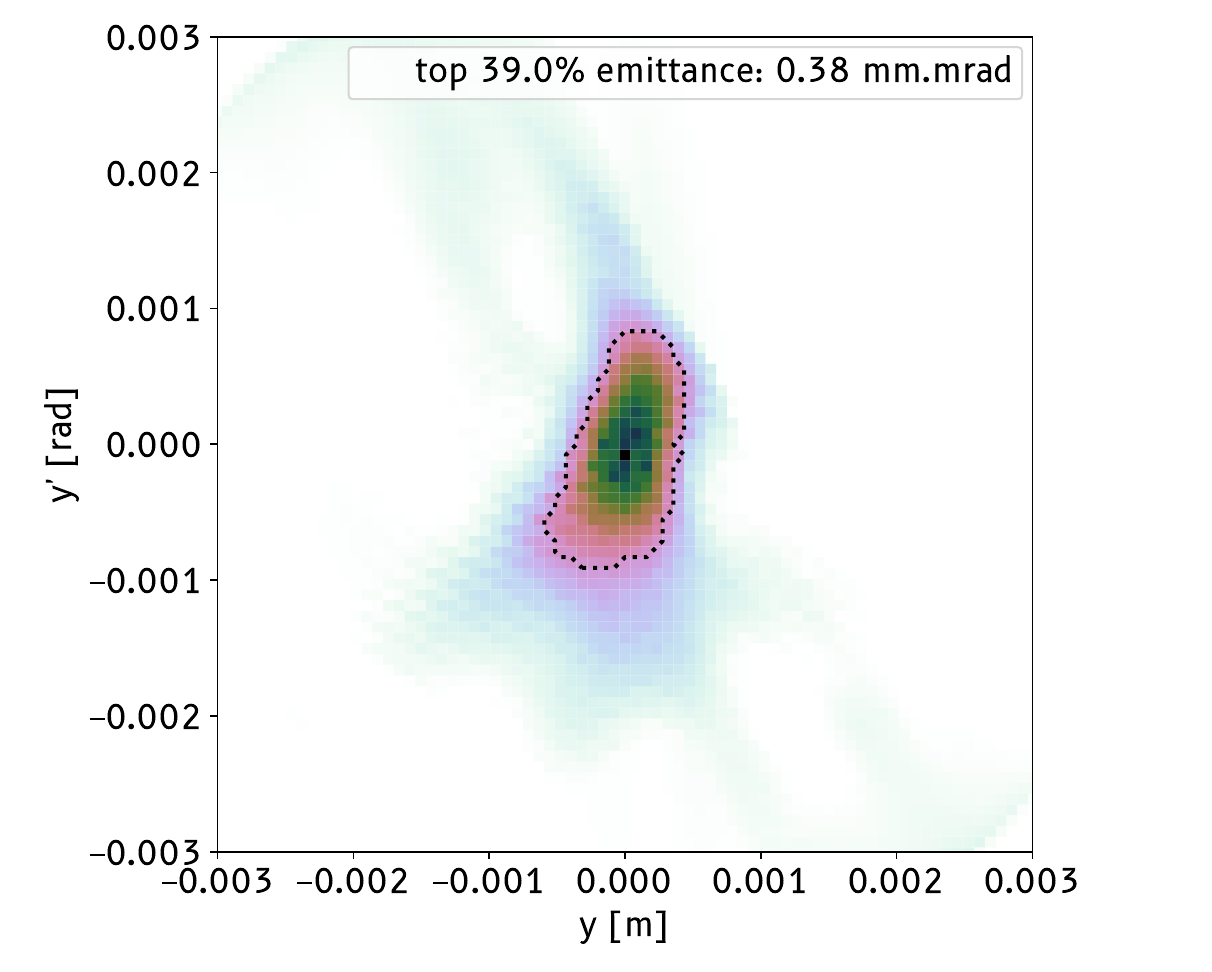}
	\caption{Example event for single-shot emittance determination. Clockwise from top-left: input spectrometer image; fit to vertical beam size; tomographically reconstructed phase space. Note: 39\% of the volume of a two dimensional Gaussian is enclosed inside a one sigma contour. The 39\% volume-enclosing contour is shown as a dotted black line.}
	\label{fig:example}
\end{figure}

Both multishot and single shot quadrupole scans can be analysed by fitting upstream beam parameters ($\sigma_{y_0}^2$, $\sigma_{y_0y_0'}$, and $\sigma_{y_0'}^2$) to the measured vertical size squared ($\sigma_y^2$) with a function of the vertical plane transport matrix elements $R_{11}$ and $R_{12}$ (where $R_{11}$ is the coupling between $y_0$ and $y$, and $R_{12}$ the coupling between $y_0'$ and $y$):
\begin{align}
	\sigma_y^2 = \sigma_{y_0}^2 R_{11}^2 + \sigma_{y_0y_0'} R_{11}R_{12} + \sigma_{y_0'}^2 R_{12}^2\label{eqn:beamsize}
\end{align}
The difference between the two methods arising from $R$ being a function of $k$ in the multishot case ($k$ being the normalized quadrupole strength), and the energy $E$ being a free choice (the size should be measured at the same position on the screen for each shot); and $R$ being a function of $E$ in the single shot case, since the quadrupole strength is not varied explicitly, and the measurement positions are dictated by the available range produced by the energy spread of the bunch. Fitting with equation \ref{eqn:beamsize} requires that the resolution be included explicitly, either by adjusting the measured size by subtracting in quadrature a PSF width, which assumes the PSF itself is Gaussian, or more generally by performing a deconvolution directly on the beam images, which allows for arbitrary forms of the PSF \cite{Lucy1974}. An example of a fit to vertical beam size using this method is shown in Figure \ref{fig:example}.\par

Emittance reconstructed by either of these methods assumes the bunch is Gaussian, as we reconstruct a beam matrix and use the standard result that the geometric emittance is the square root of the determinant of this matrix. Under certain assumptions, these datasets can also be used to reconstruct the $(y, y')$ phase space tomographically using an inverse Radon transform \cite{McKee1995}, meaning that a more general measurement of the phase space area can be made (emittance in this case the area of the top 39\% divided by $\pi$, which, for Gaussian beams, is identical to the previous definition). Phase space tomography uses the transport matrix from a point before the quadrupoles to the screen to calculate an effective rotation $\theta$ of the phase space, and a scaling factor $s$ (required in order to map it to a regular Radon transform), defined as follows:
\begin{align}
	\theta &= \frac{R_{12}}{R_{11}}\\
	s &= \sqrt[]{R_{11}^2 + R_{12}^2}
\end{align}
Typically, $(x,y)$ beamspot images taken with different quadrupole strengths are then summed along an axis to produce profiles, which are then scaled and stacked into a \textit{sinogram}, upon which standard tomographic reconstruction algorithms (e.g filtered back-projection \cite[e.g.][]{Kak1979}, simultaneous algebraic reconstruction \cite{Andersen1984}) may be run to produce an image of the phase space in a particular plane (e.g. $(y,y')$ if the original summing was along $x$). In this particular case, each vertical slice of a spectrometer image is again assumed to represent the whole bunch, and taken as a $y$ profile by itself. Thus, the spectrometer images require only scaling and columnwise normalization to be considered sinograms. As the images are taken with digital cameras, it is natural to use the pixel columns as the energy slices (even if the resolution of the spectrometer imaging system is larger than one pixel), and so no further processing of the images is required. An example of a reconstructed phase space is shown in Figure \ref{fig:example}.

\section{Results}

\begin{figure}
	\centering
	\includegraphics[width=0.49\textwidth]{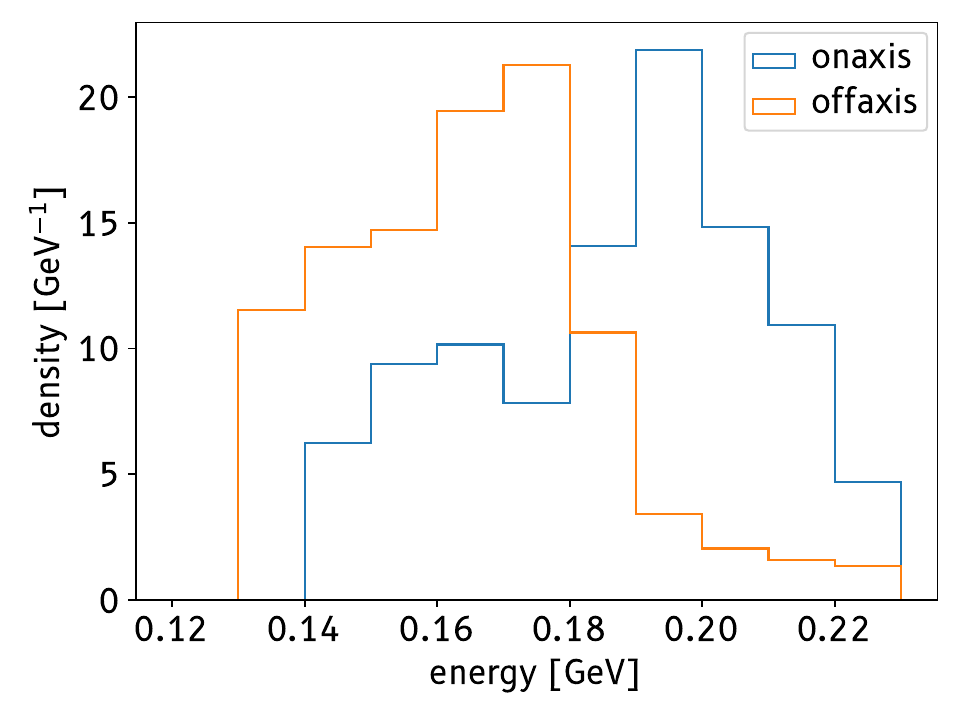}
	\includegraphics[width=0.49\textwidth]{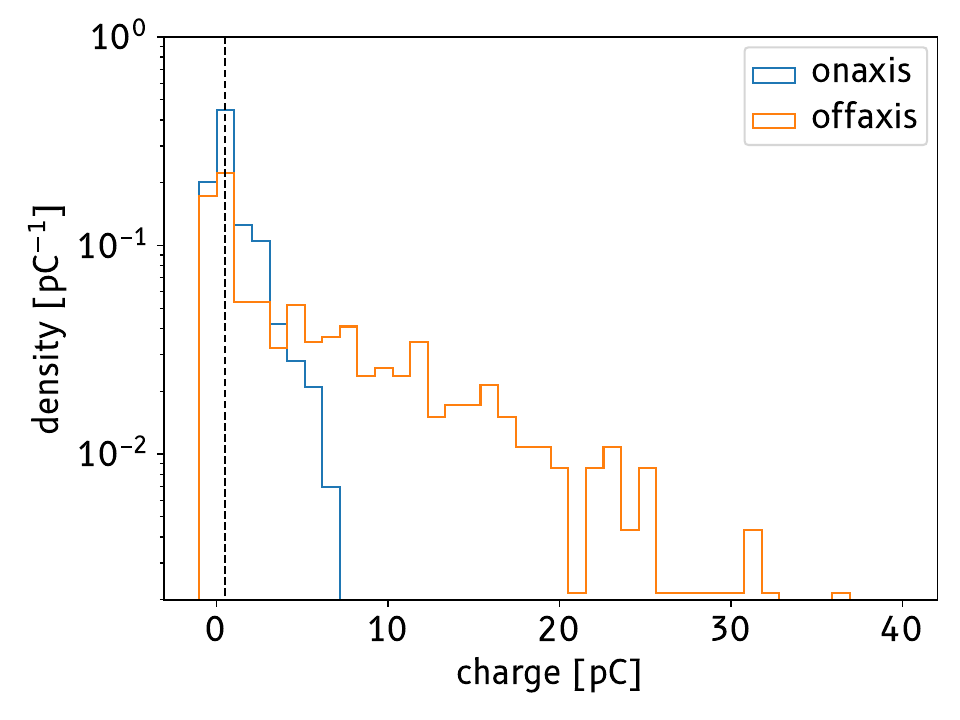}
	\caption{Distribution of peak energy of capture bunches (left) and of captured charge, showing wide variation (right). The background charge cut below which events are rejected is shown as a vertical dashed line.}
	\label{fig:energycharge}
\end{figure}
Figures \ref{fig:energycharge} shows the energy and charge capture distributions for the events in the dataset. Energy gain is around 200 MeV, but charge capture varies considerably. Event selection is limited to removing those events below a fixed charge threshold of \SI{0.5}{\pico\coulomb}. Further event selection using classification by energy distribution profile using a clustering algorithm to group similar events together could strengthen the assumption that, for a multishot quadrupole scan, the beam parameters are not varying from shot-to-shot, and for single shot measurements, allow a more like-for-like comparison. However, the small dataset sizes used here resulted in clusters with populations too low for analysis. Note that the working point is very far from the AWAKE baseline of plasma density of \SI{7e14}{\per\centi\metre\cubed} and proton bunch population of $3\times10^{11}$, as data acquisition was optimized for charge capture reliability and energy gain stability, owing to the short data acquisition period available for the measurement. Energy gain is therefore not expected to be near the \SI{2}{\giga\electronvolt} previous reported in \cite{Adli2018}.\par

\begin{figure}
	\centering
	\includegraphics[width=0.49\textwidth]{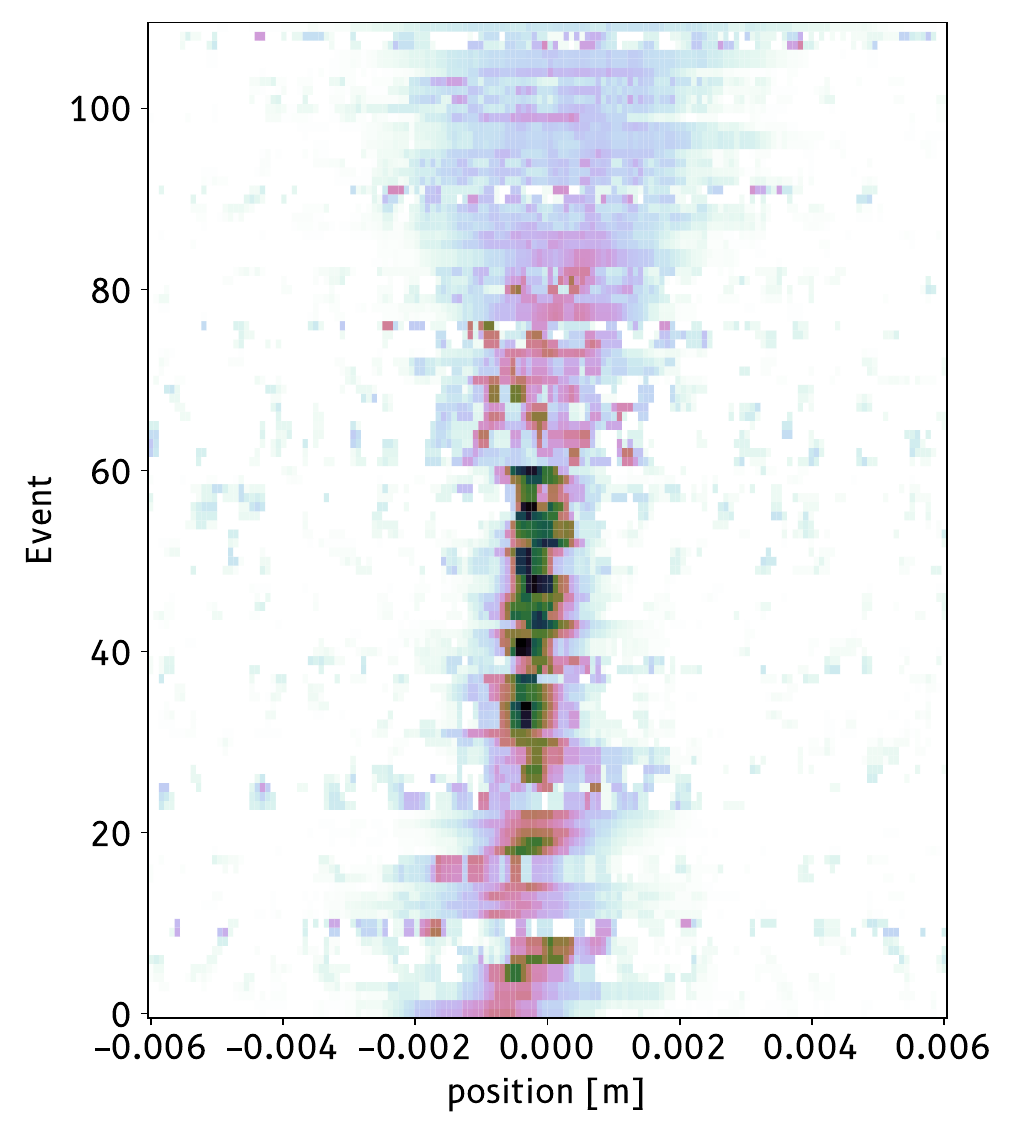}
	\caption{Waterfall plot of one vertical energy slice, sorted by quadrupole strength. Each slice is normalized by the area to aid visibility; as a result of energy variation between shots, some events have no intensity in this slice. Plotted relative to the vertical centre of the screen.}
	\label{fig:waterfall}
\end{figure}

Figure \ref{fig:waterfall} shows a waterfall plot of charge-normalized, quadrupole-strength-sorted spectrometer image slices. This serves as a visual check that the beam parameters are not changing so extremely from shot to shot as to distort the expected behaviour in a quadrupole scan. It also illustrates a particular energy slice from which the emittance can be determined using the multishot technique.

\begin{figure}
	\centering
	\includegraphics[width=0.49\textwidth]{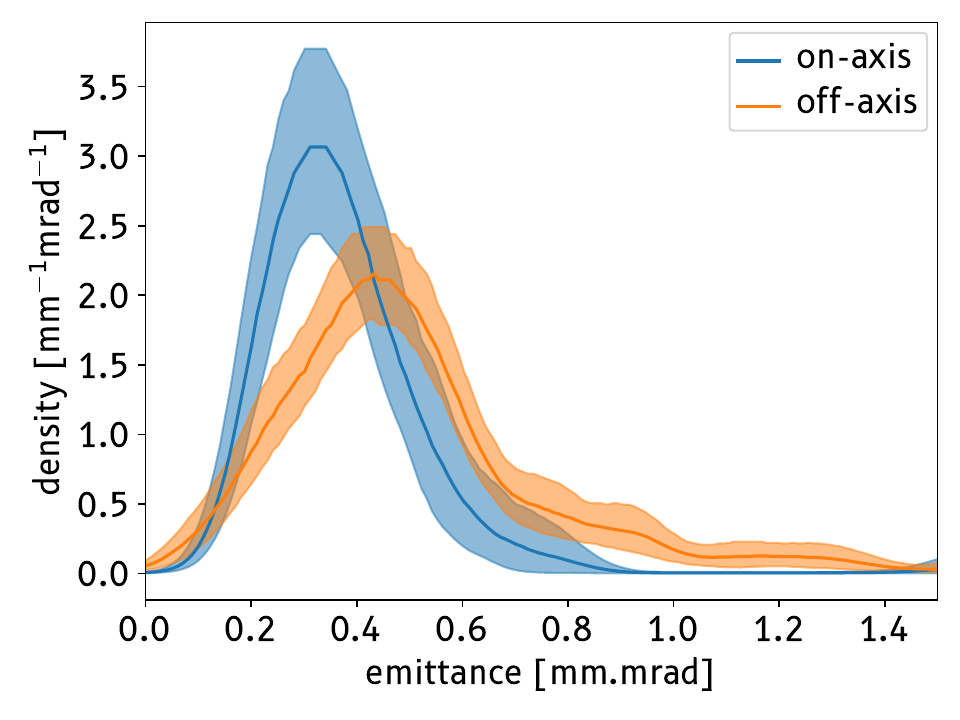}
	\includegraphics[width=0.49\textwidth]{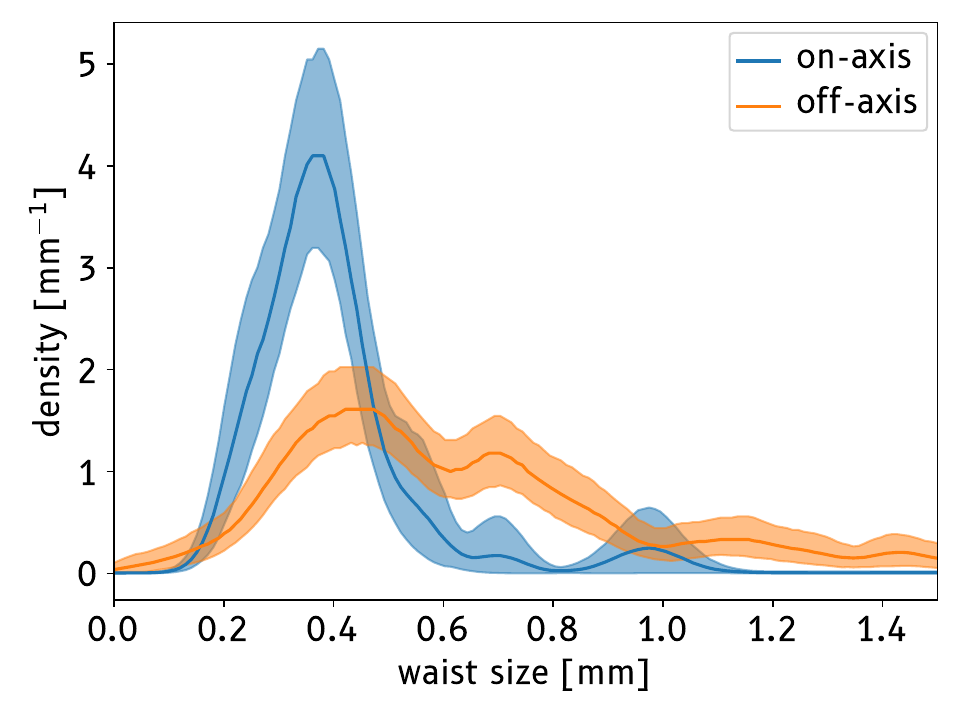}
	\includegraphics[width=0.49\textwidth]{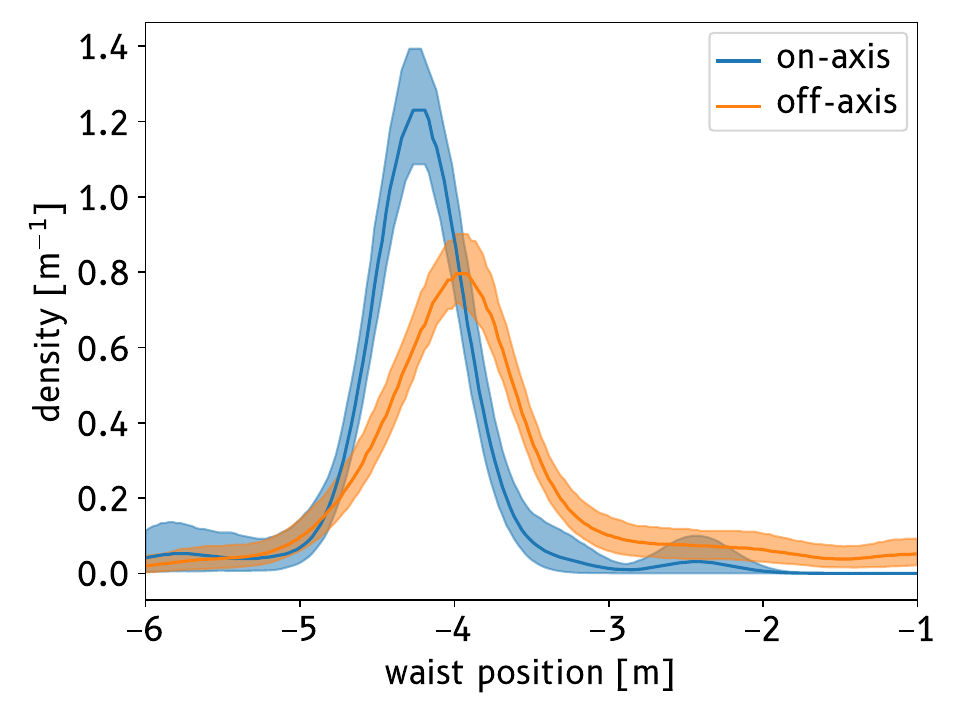}
	\includegraphics[width=0.49\textwidth]{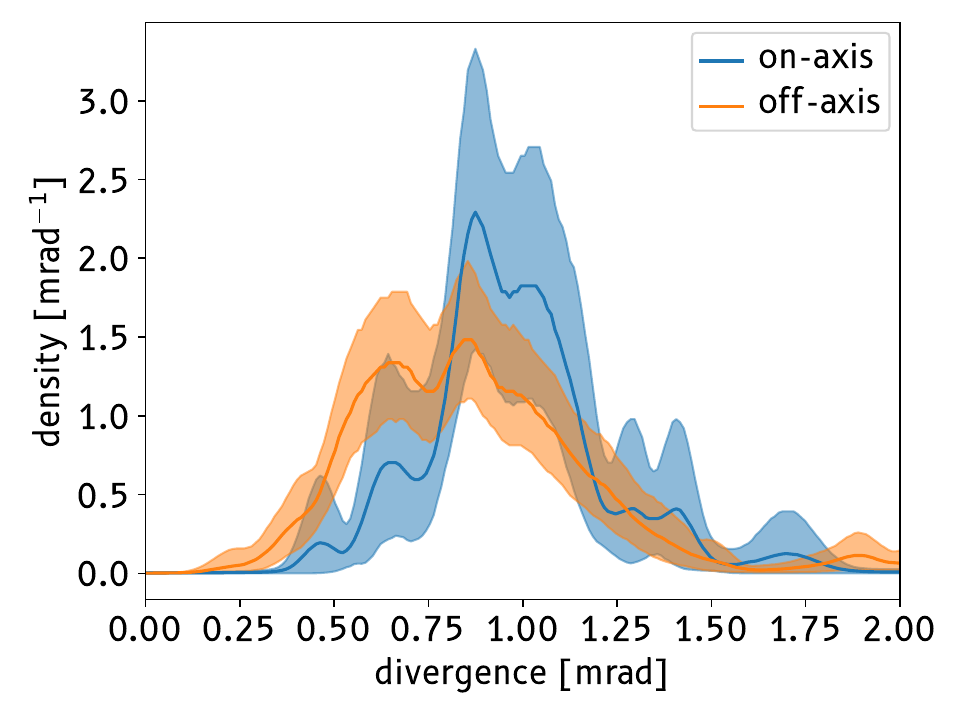}
	\caption{Clockwise from top left, emittance, beam size at waist, divergence, and waist position (relative to quadrupole position) for on- and off-axis datasets, as determined by single-shot beam size measurements.}
	\label{fig:ss}
\end{figure}

\begin{figure}
	\centering
	\includegraphics[width=0.49\textwidth]{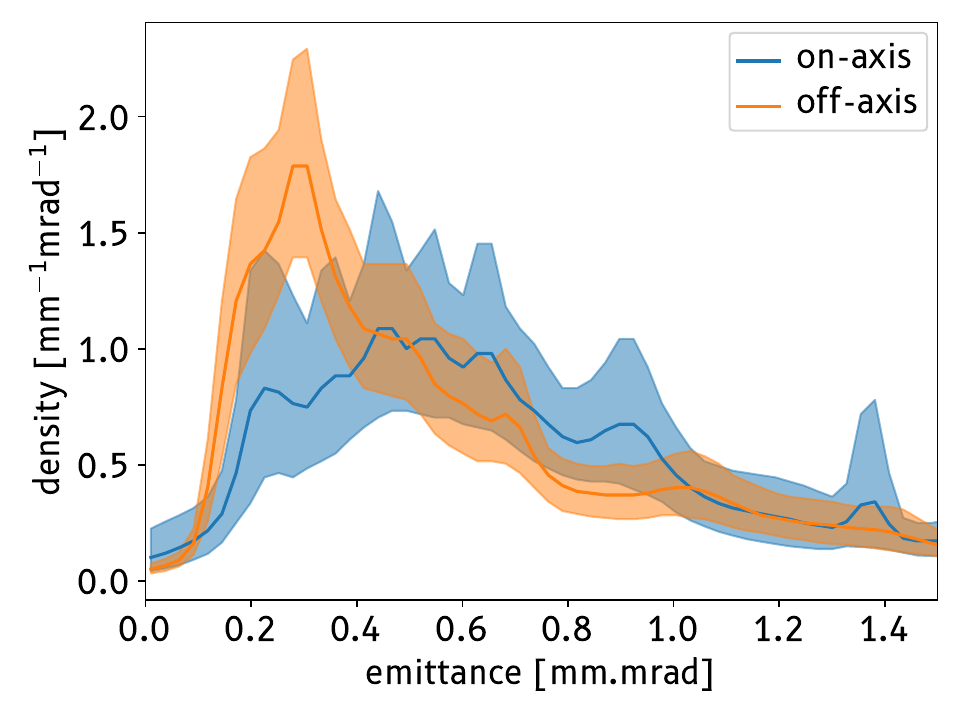}
	\includegraphics[width=0.49\textwidth]{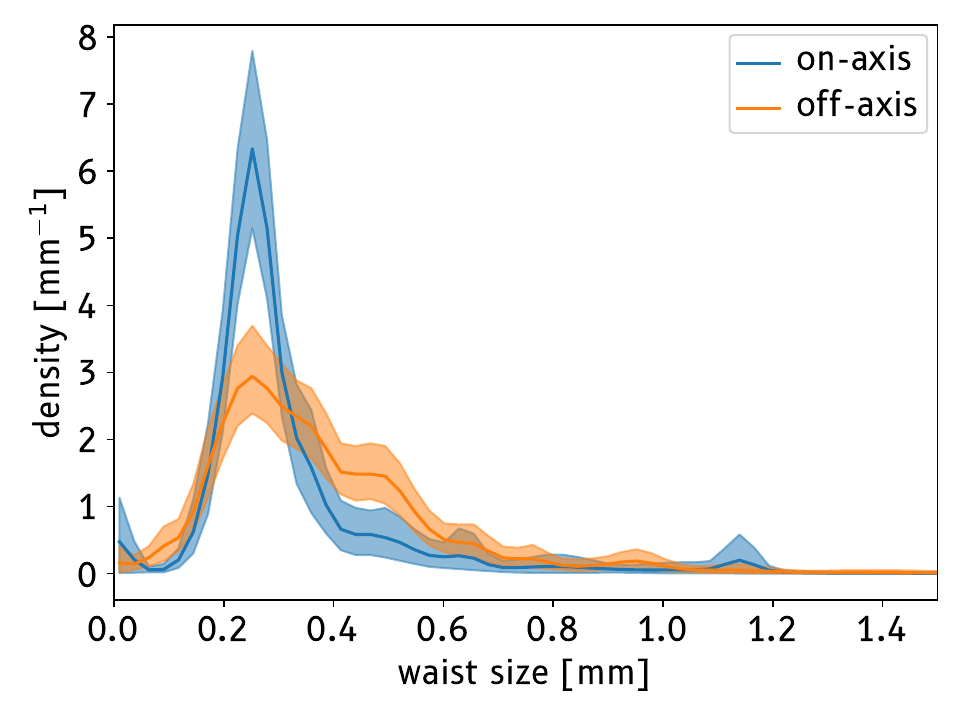}
	\includegraphics[width=0.49\textwidth]{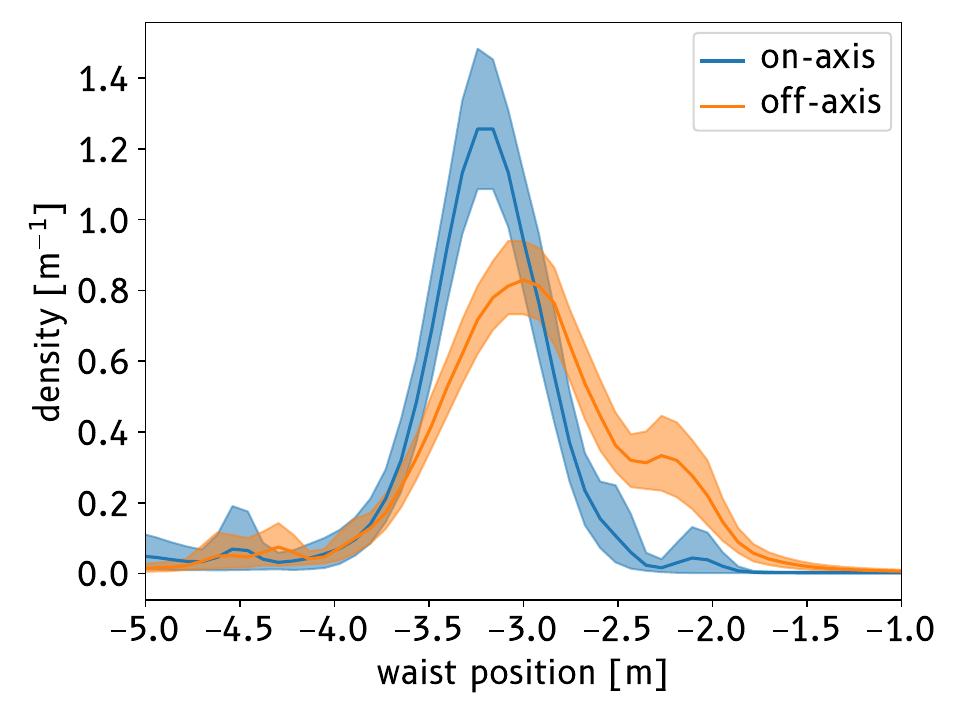}
	\includegraphics[width=0.49\textwidth]{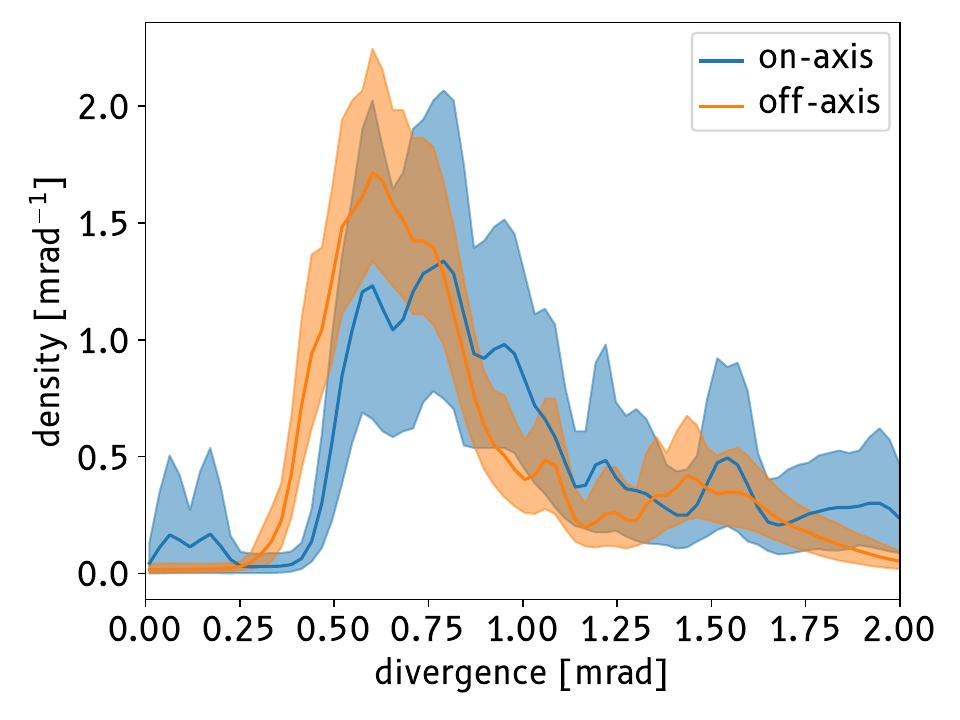}
	\caption{Clockwise from top left, emittance, beam size at waist, divergence, and waist position (relative to quadrupole position) for on- and off-axis datasets, as determined by single-shot phase-space tomography measurements.}
	\label{fig:pst}
\end{figure}

\begin{figure}
	\centering
	\includegraphics[width=0.49\textwidth]{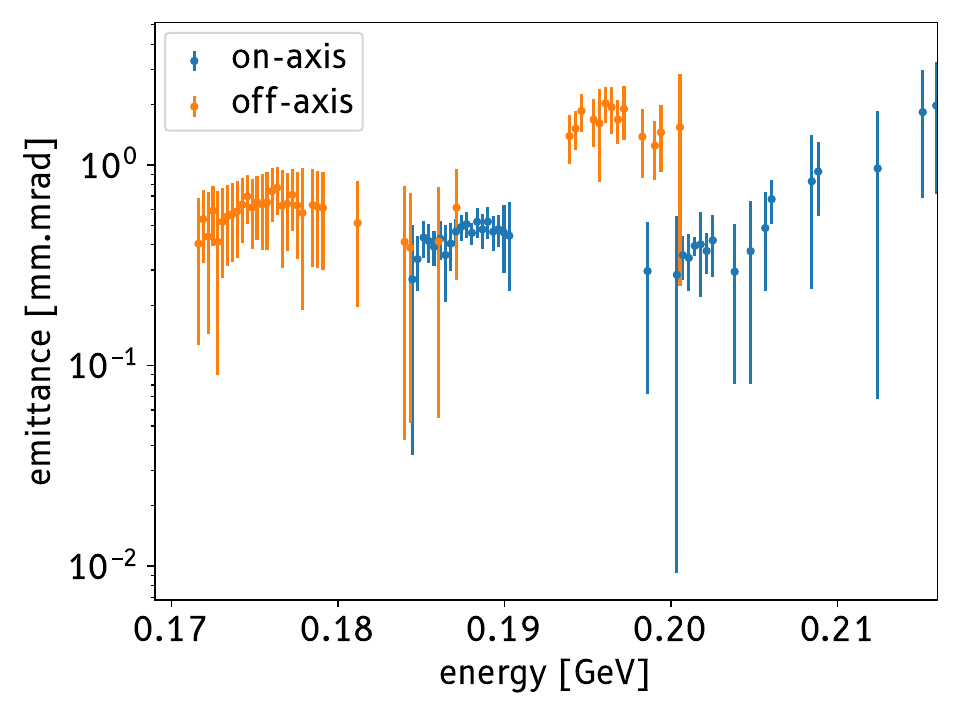}
	\includegraphics[width=0.49\textwidth]{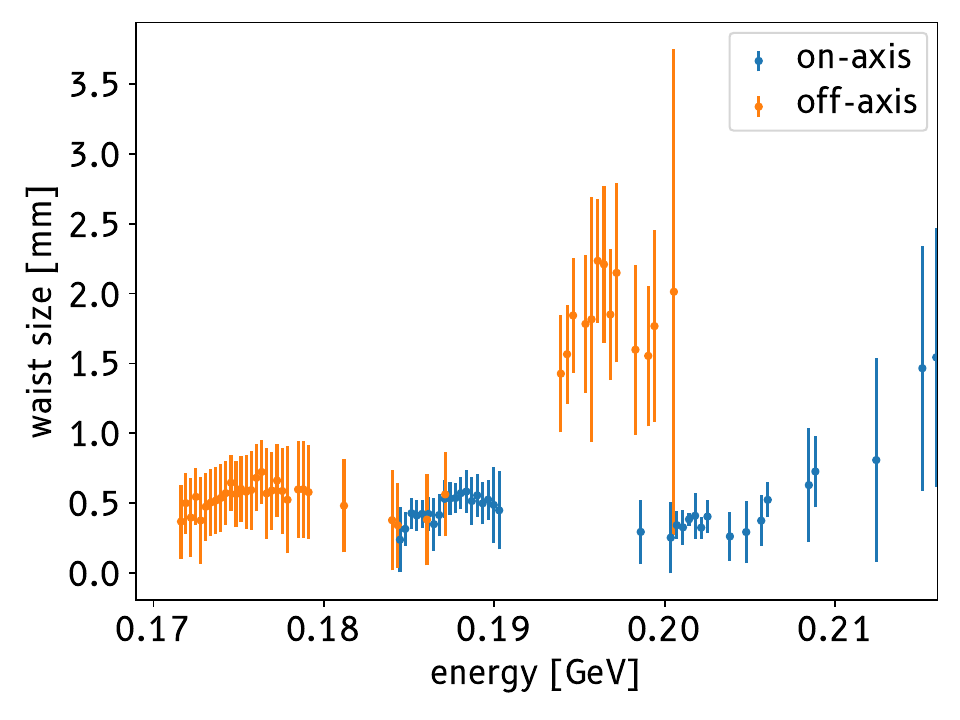}
	\includegraphics[width=0.49\textwidth]{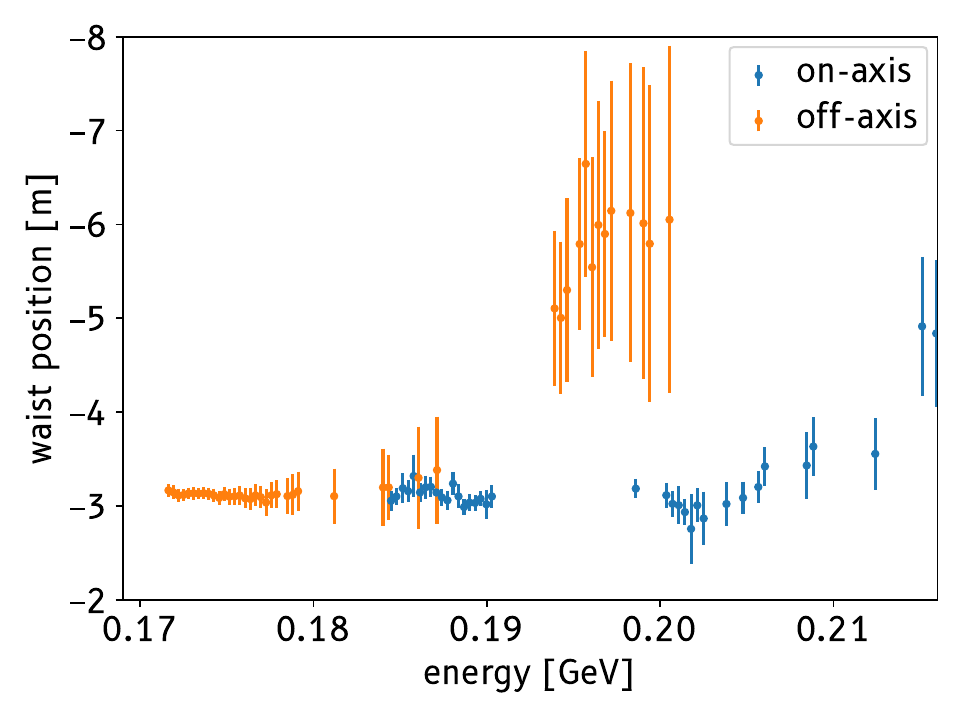}
	\includegraphics[width=0.49\textwidth]{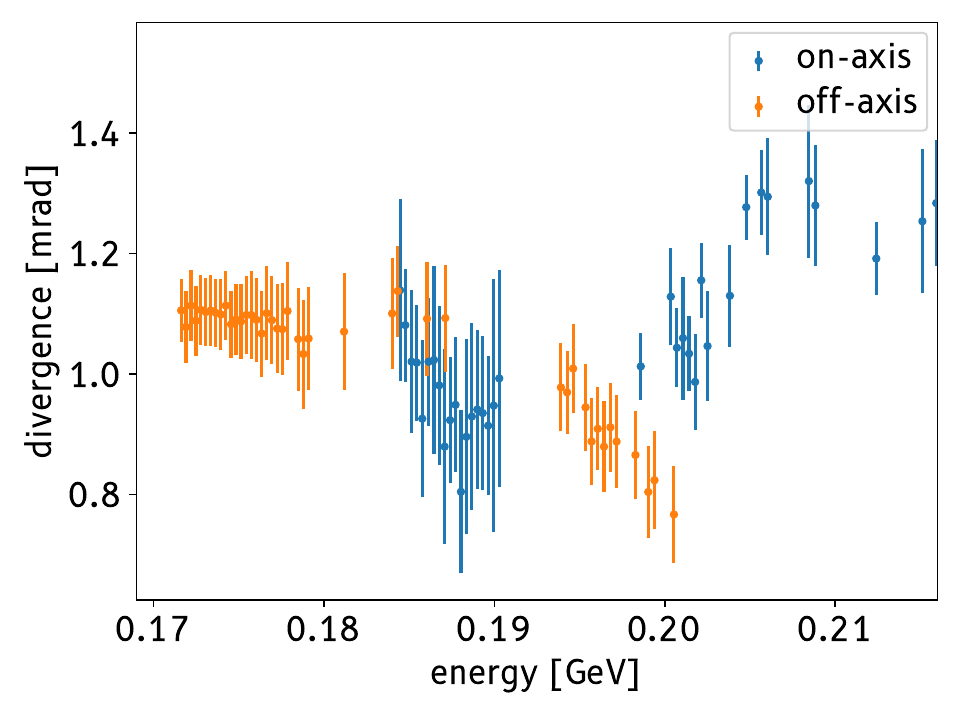}
	\caption{Clockwise from top left, emittance, beam size at waist, divergence, and waist position (relative to plasma exit iris), as determined by multishot quadrupole scan measurements.}
	\label{fig:ms}
\end{figure}

Figure \ref{fig:ss} shows geometric emittance, waist size, and waist position distributions for the single shot beam size measurement method. The on- and off-axis dataset size in each case is 130 and 400 events, respectively. To include the calculated uncertainties on the fitted parameters, raw data histograms are transformed into kernel density estimates, using a Gaussian kernel for each point, with width given by the uncertainty. The systematic uncertainties are dominated by the dipole current ripple, which has a particularly pronounced effect on the determination of the beam waist location. Two-$\sigma$ uncertainty bands around the distribution are generated from bootstrap sampling with replacement to produce an empirical cumulative distribution function for each bin, with $\pm 2\sigma$ approximated from the bin count range between the 2.5\% and 97.5\% distribution points. The accumulation in the histograms at preferred values appears to support the assumption that the beam parameters do not vary greatly from shot-to-shot, which is important for validation of the multishot technique.\par

Also derived from single-shot measurements are the phase space tomography results shown in Figure \ref{fig:pst}. In this case, the emittance is defined as the area occupied (see Figure \ref{fig:example}) by the top 39\% of pixels by intensity, divided by $\pi$. This measure gives the area inside a one $\sigma$ contour for a Gaussian phase space (division by $\pi$ is so the result then matches that given by the square root of the determinant of a Gaussian beam matrix). Peak values for the beam parameters appear commensurate with those determined by beam size measurements, if the data are drawn from a single population so that the histogram width can be interpreted as an uncertainty on the central value (see Tables \ref{tab:sum1} and \ref{tab:sum2}).\par

\begin{table}
	\centering
	\caption{Summary of beam parameters extracted by the three methods, for the on-axis dataset. Values in parentheses are 1.48 times the median absolute deviation, used as a robust estimate of the standard deviation.}
	\label{tab:sum1}
	\begin{tabular}{r|S[table-format=-1.2(2)]S[table-format=-1.2(2)]S[table-format=-1.2(2)]}
		\hline\hline
		{Quantity$\downarrow$/Method$\rightarrow$}	& {Beam size}	& {Tomography}	& {Quadrupole scan}\\\hline
		Waist size (mm)								& 0.38\pm0.07	& 0.26\pm0.02	& 0.41\pm0.12	\\
		Waist position (m)							& -4.25\pm0.14	& -3.21\pm0.13	& -3.12\pm0.09	\\
		Divergence (mrad)							& 1.00\pm0.15	& 0.94\pm0.30	& 1.07\pm0.10	\\
		Emittance (mm mrad)							& 0.37\pm0.10	& 0.66\pm0.28	& 0.42\pm0.10	\\\hline\hline
	\end{tabular}
\end{table}
\begin{table}
	\centering
	\caption{As Table \ref{tab:sum1}, but for the off-axis dataset.}
	\label{tab:sum2}
	\begin{tabular}{r|S[table-format=-1.2(2)]S[table-format=-1.2(2)]S[table-format=-1.2(2)]}
		\hline\hline
		{Quantity$\downarrow$/Method$\rightarrow$}	& {Beam size}	& {Tomography}	& {Quadrupole scan}\\\hline
		Waist size (mm)								& 0.60\pm0.21	& 0.33\pm0.09	& 0.47\pm0.10	\\
		Waist position (m)							& -3.92\pm0.32	& -2.97\pm0.28	& -3.18\pm0.07	\\
		Divergence (mrad)							& 0.89\pm0.24	& 0.74\pm0.20	& 1.10\pm0.01	\\
		Emittance (mm mrad)							& 0.49\pm0.15	& 0.50\pm0.21	& 0.52\pm0.10	\\\hline\hline
	\end{tabular}
\end{table}

Figure \ref{fig:ms} shows the same results, acquired using the traditional quadrupole scan technique. As the spectrometer disperses horizontally by energy, these quadrupole scans can be performed at many energies simultaneously, and so the results are presented as a function of energy. Immediately it can be seen that for a range of energies (\SI{0.18}{\giga\electronvolt}--\SI{0.21}{\giga\electronvolt} for on-axis; \SI{0.17}{\giga\electronvolt}--\SI{0.19}{\giga\electronvolt} for off-axis) the assumption that the beam parameters do not vary with energy is valid too. In this region, the beam parameter values agree well with the results acquired by single-shot beam size measurements.

\subsection{Resolution and validity}
\begin{figure}
	\centering
	\includegraphics[width=0.49\textwidth]{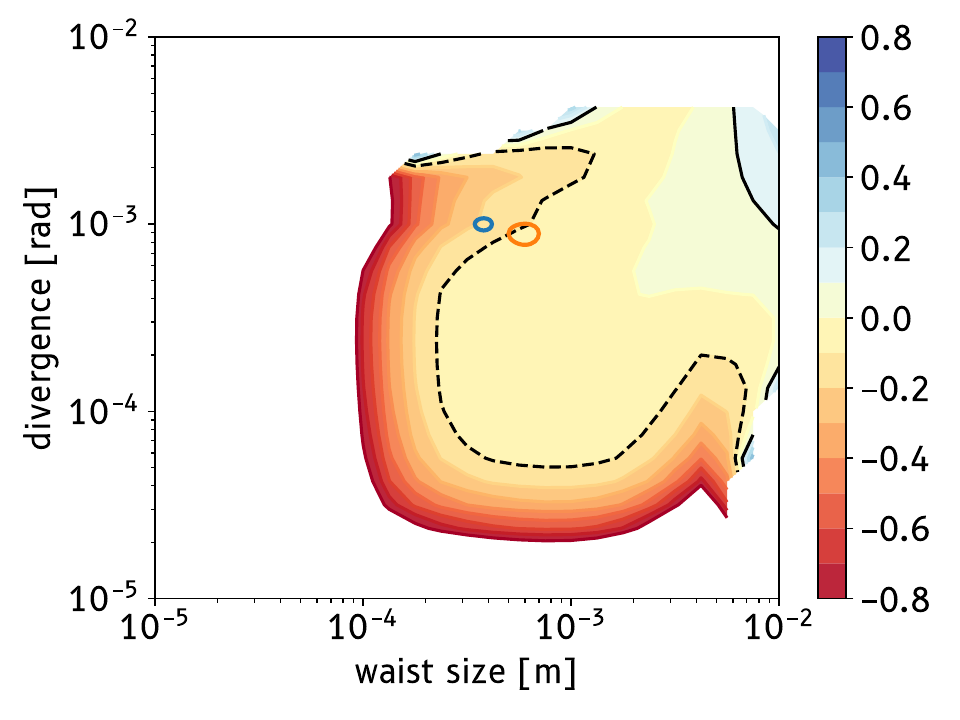}
	\includegraphics[width=0.49\textwidth]{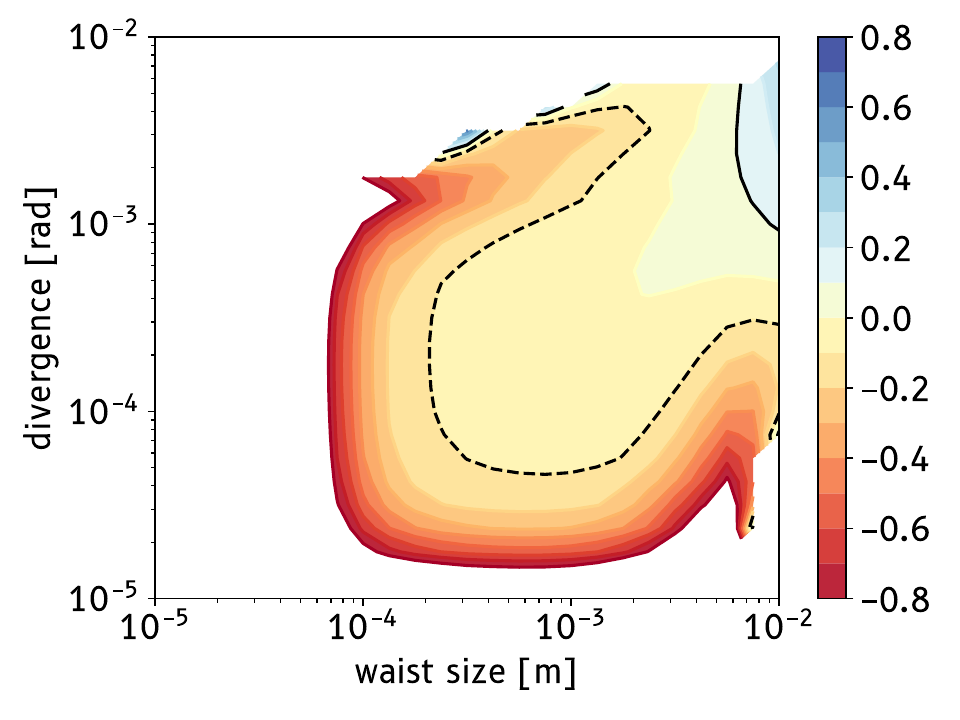}
	\includegraphics[width=0.49\textwidth]{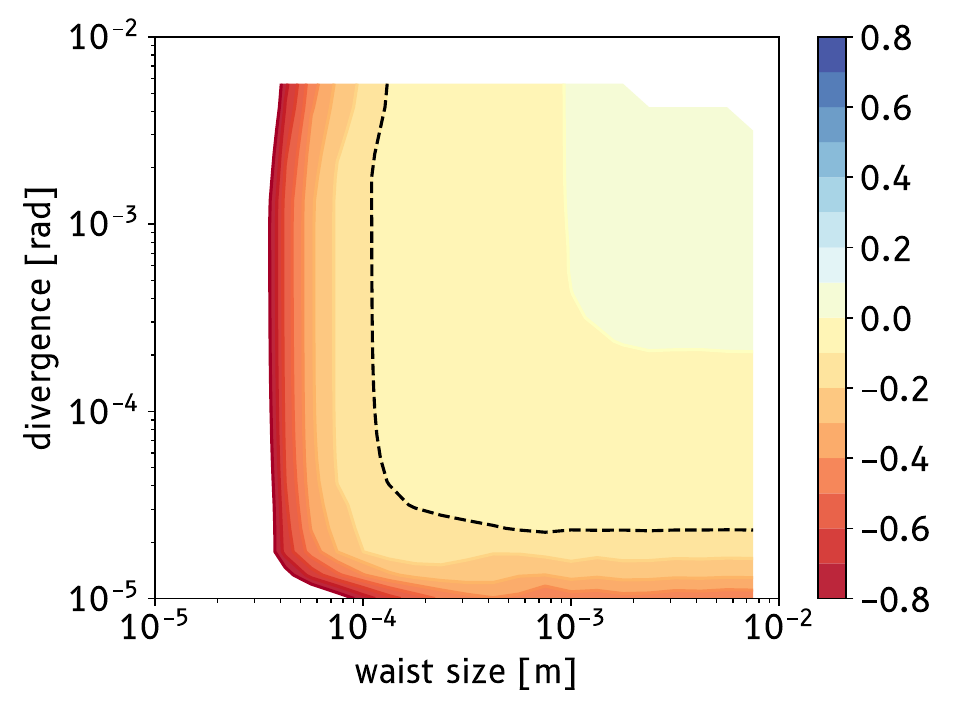}
	\includegraphics[width=0.49\textwidth]{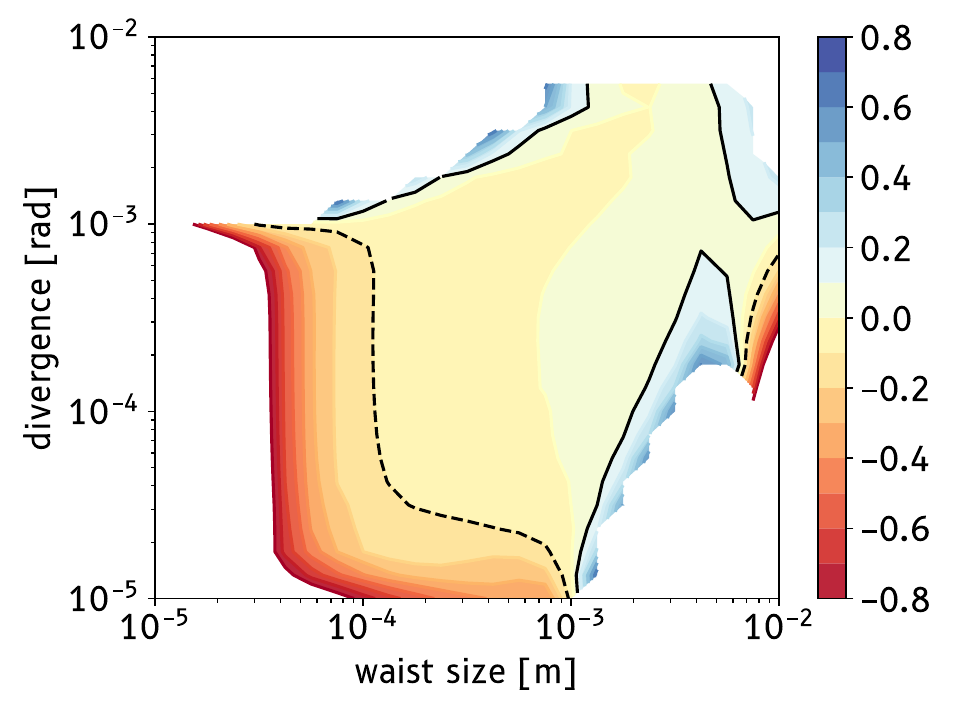}
	\caption{Emittance validity maps for different simulated parameters, showing the relative difference between input emittance and that reconstructed by fitting. Clockwise from top left: optical resolution of \SI{200}{\micro\metre} with  \SI{80}{\micro\metre} pixels; improved optical resolution (\SI{10}{\micro\metre}) with \SI{80}{\micro\metre} pixels; optical resolution (\SI{10}{\micro\metre}) with \SI{40}{\micro\metre} pixels; and optical resolution (\SI{10}{\micro\metre}) with \SI{80}{\micro\metre} pixels, but without the effects of emittance in the dispersed plane. The top left map represents the experimental conditions, and so has ellipses representing the location and scale of the on- and off-axis results (in blue and orange, respectively).}
	\label{fig:valid}
\end{figure}

The fitted beam parameters depend, to a certain degree, on the estimate of the optical resolution of the spectrometer. This has been studied, and because the dominant contribution from resolution comes from the imaged waist where the beam is smallest, and assuming Gaussian beams and PSFs, the effect is to limit the minimum measurable beam size and divergence, and that these quantities are generally linear in resolution. The first of these observations leads naturally to the question of whether the measured emittance in the above figures is correct, or limited by the resolution---that is, how valid the measurement really is. This was studied in simulation; the top-left pane of Figure \ref{fig:valid} shows the relative difference $z$ between reconstructed emittance (using the beam size method) and the simulation input emittance, as a function of beam waist size and divergence, that is: 
\begin{align}
	z(\sigma_y^{in}, \sigma_{y'}^{in}) = \frac{\epsilon_y^{in} - \epsilon_y^{fit}}{\epsilon_y^{in}}
\end{align}
where $^{in}$ and $^{fit}$ refer to the model input and fit result respectively (and e.g. $\sigma_y^{fit}$ is therefore a function of $\sigma_y^{in}$). For simplicity, the input beam is symmetric in $x$ and $y$ (both in terms of beam parameters and PSF), and both beams and PSFs are Gaussian. As the beam size method only considers the $y$ plane, it might therefore be expected that the reconstruction would fail at large emittances. This assertion has also been examined in simulation using a beam with fixed, small $x$ waist size and divergence.
Also shown in Figure \ref{fig:valid} are similar maps for different optical and experimental conditions. This shows that even dramatically reducing the resolution width has minimal impact if the camera pixel size (in the object plane) remains as large as it is. This might inform future AWAKE spectrometer optical design, where very small geometric emittances ($\sim \SI{100}{\pico\metre\radian}$), resulting from large energy gain and normalized emittance preservation under this acceleration, might be expected. These results depend on deconvolution of the spectrometer images from the point spread function, which requires good knowledge of the PSF.\par

\begin{figure}
	\centering
	\includegraphics[width=0.49\textwidth]{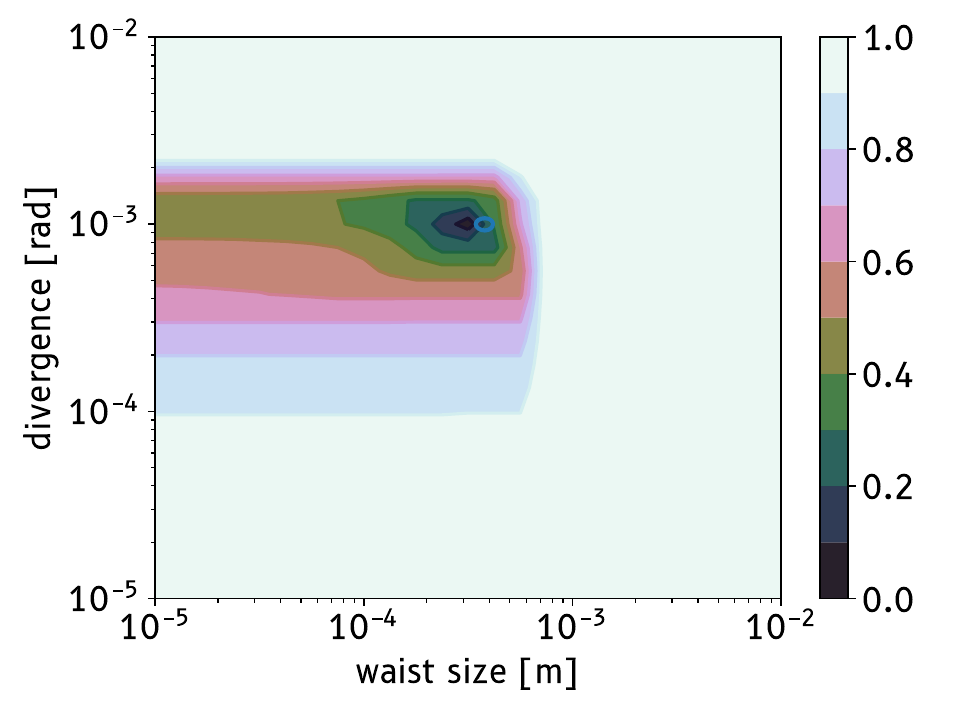}
	\includegraphics[width=0.49\textwidth]{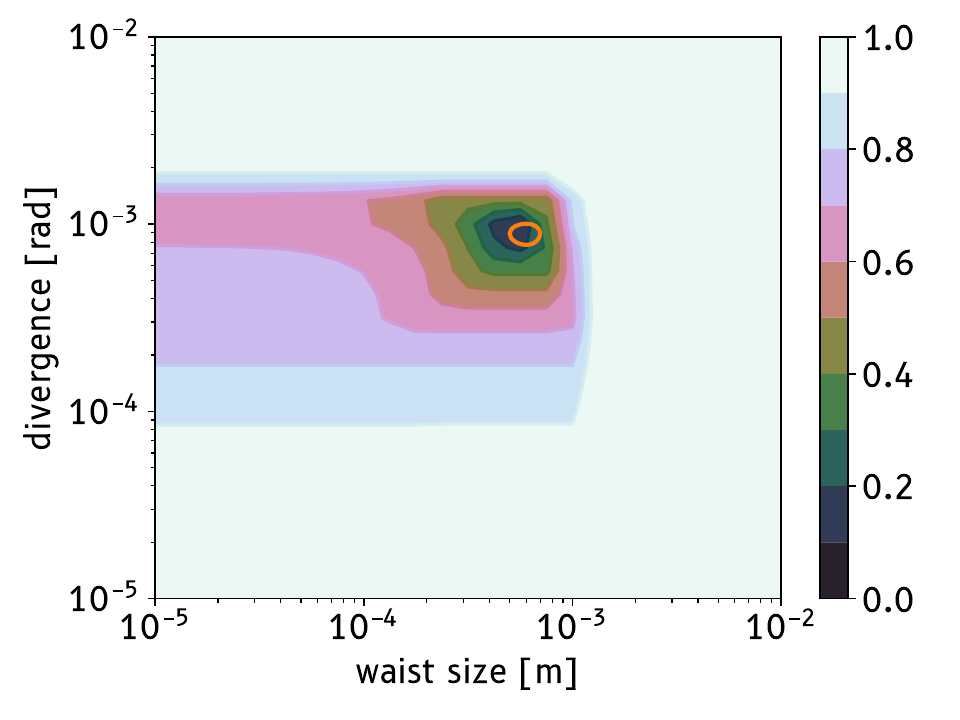}
	\caption{On- (left) and off-axis (right) maps showing the relative difference between the predicted and measured waist size and divergence, as a function of simulation input waist size and divergence. Ellipses representing the measured distributions are also shown.}
	\label{fig:valid2}
\end{figure}

Although we have shown that the experimentally determined beam parameters lie in a region of the instrument acceptance phase space where they might be expected to be well reproduced, it could be argued that the observed values simply arise from very badly reproduced input parameters that lie far outside the valid region. By using the simulation results to calculate the difference between the predicted and measured beam parameters $z$ (as in Figure \ref{fig:valid2}) as:
\begin{align}
	z(\sigma_y^{in}, \sigma_{y'}^{in}) = \frac{1}{n}\sum_{k=0}^{n} \neg\left(\left|\frac{\sigma_y^{fit} - \sigma_y}{\sigma_y}\right| < \frac{k}{n}\right) \lor \neg\left(\left|\frac{\sigma_{y'}^{fit} - \sigma_{y'}}{\sigma_{y'}}\right| < \frac{k}{n}\right)
\end{align}
(where convergence is achieved for $n > 500$, and logical true and false are treated as one and zero, respectively) it can be demonstrated that the location in phase space which \textit{produces} the observed results is very close to the observed results. This is just a restatement of the result shown in the top-left graph of Figure \ref{fig:valid}, but centred on the measured values. Since the maps derived from simulation are not of uncertainties, but fixed differences, this could be used to provide a correction to observed beam parameter to improve the results.\par

\begin{table}
	\centering
	\caption{Estimate of the effect of changing the PSF width.}
	\label{tab:resw}
	\begin{tabular}{r|l}
		\hline\hline
		$\delta\sigma_x$		&	\SI{2.4e-3}{\milli\metre\per\micro\metre}\\
		$\delta\sigma_{y'}$		&	\SI{1.9e-3}{\milli\radian\per\micro\metre}\\
		$\delta y_w$			&	\SI{3.5}{\milli\metre\per\micro\metre}\\
		$\delta\epsilon_y$		&	\SI{2.9e-3}{\milli\metre\milli\radian\per\micro\metre}\\\hline\hline
	\end{tabular}
\end{table}

These results require good knowledge of the PSF of the system, and assume it to be spatially invariant. In-situ measurements of the line-spread function were made using images of a USAF 1951 resolution test chart attached to the front of the scintillator screen. This was found to be well approximated by a Gaussian with a width of nearly \SI{100}{\micro\metre}. The remaining contribution to the resolution, arising from the electron bunches passing through the aluminium vacuum window and scintillating screen, and from photon scattering in the screen, has been estimated from a GEANT4 \cite{Agostinelli2003,Allison2006,Allison2016} simulation, and bring the width of the PSF up to approximately \SI{200}{\micro\metre}. In this study, this value is assumed, and a symmetric Gaussian form of the PSF is used. The scale of the uncertainty introduced into the values for waist size, divergence, waist position, and emittance, respectively, from systematic uncertainty in the PSF width is shown in Table \ref{tab:resw}.

\section{Discussion and summary}

The emittance and beam parameters of the accelerated electron beam at AWAKE have been determined for the first time using a variety of methods. Generally speaking, good agreement is found between the various methods, although there are notable exceptions. The beam parameters, and basic properties of the bunches (captured charge and energy gain) form distinct populations separated by injection angle, indicating that a difference between injection conditions remains detectable after the acceleration process. Differences observed between the analysis methods may be attributed to the different assumptions each of the methods contains. In order for all the three methods to produce identical results, beam parameters must be the same at all energies in a bunch, be the same across many event, be well described by gaussian distributions, and bunches must have wide enough energy spreads for good reconstruction. If these conditions are slightly violated, we might expect to see slight disagreement between parameter values determined by these different methods. This occurs for some parameter values, for example, the off-axis quadrupole scan divergence, or the waist location determined by single-shot beam size measurements. The variations between methods are, for the most part, small, and the methods are taken as producing results in agreement, but with complementary information.\par

The beam size, divergence, and waist positions are all \textit{reasonable}, in the sense that they are on the correct scale, but the phase space area occupied by the accelerated beam is much larger than that of the injection beam. The injection scheme used presently leads to complex beam dynamics with a number of different processes that can affect the phase space occurring at different stages. The witness bunch will drive its own wakefield when it enters the plasma, causing the bunch to pinch \cite{Bennett1955}. The focussing field is nonlinear on the scale of the witness bunch, and the resulting phase mixing leads to an increase in the bunch emittance until a quasi-equilibrium is reached \cite{Farmer2021}. Only a fraction of the initial charge is captured, so further emittance increases may occur as charge is shed and a new equilibrium must be found. In the case of off-axis injection, the bulk transverse momentum of the bunch will also be converted to an increase in the emittance.  In addition, the bunch has to enter the plasma via a ramp region of lower density \cite{Plyushchev2018} where wakefields driven by the proton beam are defocussing \cite{Gorn2018}. These factors combined mean that the phase space area is not even preserved on injection. A similar low density exit ramp introduces a non-linear defocussing effect on the accelerated beam, forces which are not accounted for in the spectrometer optics model (and which will have an effect on the reconstructed phase space). The sources of emittance growth during injection and acceleration will be avoided in AWAKE Run 2c \cite{Gschwendtner2022} by using a witness bunch injected on-axis with a radius and emittance matched to the focussing wakefields it will generate. Injecting after the development of the self-modulation of the proton driver should allow for essentially 100\% charge capture.\par

The present measurements serve as a proof-of-principle for accelerated beam emittance measurements using the AWAKE electron spectrometer. Single-shot emittance measurement is an important capability for the AWAKE experiment going into later running stages, where emittance preservation is one of the goals. The primary challenge to translating this method to AWAKE run 2c geometry and beam parameters will come from the resolution of the optical system of the spectrometer. Currently the limit to resolution arises from the use of relatively thick scintillating screens, mounted on an aluminium vacuum window. Considerably smaller accelerated beam sizes are anticipated in future AWAKE runs \cite{Gschwendtner2022}, and although some magnification can be achieved with electron optics, subject to space and cost--complexity constraints, the switch to self-supporting thin in-vacuum scintillating screens, or indeed, optical transition radiation screens, may still be necessary. Although the validity results shown here demonstrate that the ultimate lower limit on the measurable size and divergence arises from the pixel size in the object plane, resolution should be improved wherever possible to avoid overreliance on deconvolution algorithms, which may introduce artifacts \cite[e.g.][]{Yongpan2010}, and additionally to reduce the effects of the relative uncertainty of the PSF width.

\begin{acknowledgments}
This work was supported in parts by Fundação para a Ciência e Tecnologia - Portugal (Nos.\ CERN/FIS-TEC/0017/2019, CERN/FIS-TEC/0034/2021, UIBD/50021/2020); STFC (AWAKE-UK, Cockcroft Institute core, John Adams Institute core, and UCL consolidated grants), United Kingdom; the National Research Foundation of Korea (Nos.\ NRF-2016R1A5A1013277 and NRF-2020R1A2C1010835).
M. W. acknowledges the support of DESY, Hamburg.
Support of the Wigner Datacenter Cloud facility through the Awakelaser project is acknowledged.
TRIUMF contribution is supported by NSERC of Canada.
UW Madison acknowledges support by NSF award PHY-1903316.
The AWAKE collaboration acknowledges the SPS team for their excellent proton delivery.
\end{acknowledgments}

\bibliography{awake-accel-emittance-bibliography}

\end{document}